# HST and Fundamental Physics: Working Group Report

*September 2017*

*Preamble: The HST and Fundamental Physics Working Group was constituted by STScI Director Ken Sembach to identify ways that the Hubble Space Telescope could enable advances in fundamental physics.*

*Authors: Neal Dalal, Cora Dvorkin, Jeremy Heyl, Bhuvnesh Jain (Chair), Marc Kamionkowski, Phil Marshall, David Weinberg*

*STScI Facilitator: Neill Reid*

## 1. Introduction

The Hubble Space Telescope has been the most scientifically productive telescope of our times and possibly of all times, with a stunning array of discoveries in just about every area of astronomy.  The scope of HST's influence has extended further, to impact our understanding of the fundamental laws of physics. Looking ahead, HST can do even more if fundamental physics is promoted from an auxiliary area to one of the central goals of HST's mission.

The primary aim of astronomy is to discover, learn about, and understand the plethora of objects in the Universe, from planets, exoplanets, protostars, stars of various types, galaxies, groups, galaxy clusters, and more.  Many astronomical studies focus on individual objects while others do population studies of vast numbers of related objects.  This traditional "object-oriented" astronomy can be contrasted with cosmology, which we here define to be the study of the origin, structure, contents, and evolution of the Universe as a whole.  HST has made essential contributions to this endeavor as well through, for example, measurements of the Hubble constant (that are ongoing but that date back to the Hubble Constant Key Project).

Cosmology has become intrinsically intertwined with the study of the laws of physics.  The nature of the dark matter requires new physics beyond the Standard Model of elementary-particle interactions and Einstein's general relativity, as does the nature of dark energy (whose existence was established in part through HST observations).  Theories for the origin of the primordial density perturbations that are seen in fluctuations of the CMB and that seed the growth of large-scale cosmic structures also require new physics.  New physics is further required to account for the prevalence of baryons over antibaryons in our Universe.

Over the past two decades, cosmological observations have provided strong support for a "standard model" of cosmology, known as ΛCDM, which successfully explains a wide range of phenomena, from cosmic microwave background (CMB) anisotropies imprinted at z = 1100 to large scale structure in the galaxy and matter distributions at z = 0.  Physically, ΛCDM incorporates cold dark matter, a spectrum of primordial fluctuations from inflation, a spatially flat universe, and dark energy that is constant in space and time (Λ).  A central goal of "precision cosmology" is to test ΛCDM as stringently as possible in the hope of finding new clues to the nature of dark energy, dark matter, primordial fluctuations, or the workings of gravity on cosmic scales. There are also some intriguing tensions between various cosmological measurements -- for example, that between values of the Hubble constant inferred from the CMB and from local measurements -- that may require new physics.  Progress on answering these questions requires the type of precision measurements and control of systematics typical of particle-physics experiments, rather than the discovery/description mode that drives much of object-oriented astronomy.

Astronomical measurements can impact fundamental physics in several other ways.  For example, the study of astrophysical compact objects can be used to test the laws of physics in physical regimes inaccessible in the laboratory, and they can be used to seek spatial or time variation of fundamental constants.  An array of measurements, mostly based upon gravitational lensing, can be used to test gravity and the nature of dark matter in a variety of environments.

The unique observational capabilities of HST that have made it such a valuable discovery tool also make it a unique tool, among an array of others (particle accelerators, direct dark-matter searches, galaxy surveys, CMB, neutrino experiments and others), in the quest for new physics.  We thus advocate that HST be viewed as a multipurpose facility, one that can be used for fundamental-physics measurements, as well as for object-oriented astronomy. We describe below topics on which HST resources can impact in the coming years; these are meant to be illustrative, since our main recommendation is a new HST program that aims to address fundamental physics questions.

# 2. Hubble's Constant

The Hubble constant $H_0$ is the oldest observable of modern cosmology, with the first (badly inaccurate) estimate already appearing in Hubble's 1929 paper on the expanding universe.  The HST Key Project (Freedman et al. 2001) achieved a measurement of $H_0$ with 10% uncertainty, a dramatic improvement over the factor-of-two range of values that prevailed before HST.  There have been advances on many fronts since the completion of the Key Project, with multiple studies in recent years achieving estimated errors better than 5% (e.g., Riess et al. 2011; Freedman et al. 2012; Sorce et al. 2012; Riess et al. 2016; Bonvin et al. 2017).  Notably, Riess et al. (2016, hereafter R16) use Cepheid calibration of Type Ia supernova (SNIa) distance indicators to infer $H_0$ = 73.2 ± 1.7 km s$^{-1}$ Mpc$^{-1}$, with a 2.4% error bar including both statistical and identified systematic uncertainties.

As the measurements themselves have improved, the stakes for an accurate and precise determination of $H_0$ have gone up, thanks to the sharpening of other cosmological observations.  Within the 6-parameter ΛCDM model, CMB measurements from the Planck satellite imply a Hubble constant $H_0$ = 67.8 ± 0.9 km s$^{-1}$ Mpc$^{-1}$.  With Planck data alone this inference is sensitive to the assumptions of a flat universe and a cosmological constant, but when data from baryon acoustic oscillations (BAO) and the SNIa distance-redshift relation are included the prediction becomes more robust, holding over a wide range of dark energy models.  The tension between the "cosmological" values of $H_0$ and the direct distance-ladder measurements by R16 and other groups has become one of the central issues of contemporary cosmology.  Resolving this tension may require new fundamental physics, such as unexpectedly rapid cosmic acceleration at very low redshift, or additional relativistic species that rescale the "cosmic ruler" imprinted in CMB anisotropies and BAO.  Alternatively, the current tension could be an unlucky statistical fluke, or a result of unrecognized systematics in one or more of the measurements that leads to it.

Whatever the eventual resolution of this conundrum, precise measurement of $H_0$ is a powerful complement to probes of cosmic expansion at higher redshifts and a crucial goal for fundamental physics with HST.  Measurements at the 1-2% level would put direct $H_0$ determinations on similar footing to other contemporary dark energy probes, enough to significantly improve the "figure-of-merit" return from dark energy experiments (see Weinberg et al. 2013, fig. 48).  Extrapolating current results suggests that this level of precision is within reach of ambitious but achievable HST observing programs.  Precision improvements are important, but demonstrating control of systematic uncertainties is even more important.  For these measurements to be cosmologically valuable, they must be robust enough that a discrepancy with expectations will be taken seriously as a potential indicator of new physics. There are at least three promising lines of investigation.

## 2.1 Cepheid calibration of SNIa distances

Cepheid variables are pulsating post-main sequence stars with a direct relationship between their period and luminosity.  The most precise of the current distance-ladder estimates of $H_0$ use Cepheid variable stars to calibrate the distances to nearby star-forming galaxies that have hosted well observed SNIa.  These observations calibrate the zero-point of the relation between SNIa peak luminosity and light curve shape, so that SNIa in more distant galaxies can be used to measure $H_0$ from systems for which peculiar velocity is a small correction.  The calibration of the Cepheid period-luminosity relation itself is one source of uncertainty, but that has been reduced with parallax measurements of Milky Way Cepheids from HST, and it will be further reduced by Cepheid parallaxes from the Gaia mission.  The largest source of uncertainty at present is the relatively small number of SNIa hosts (~20) with Cepheid distances.  Future observations could plausibly expand this sample to ~50 hosts out to distances of 50 Mpc.  In addition to improving statistics, future observations should aim to further improve the Cepheid period-luminosity calibration at the longer periods used for more distant SNIa hosts, to control systematics associated with photometric biases in crowded fields and the metallicity dependence of the period-luminosity relation, and to improve the SNIa relative distance measurements with high-resolution near-IR imaging.  Reasonable projections suggest that an ambitious program could achieve 1% precision on $H_0$, but this is a demanding goal that requires tight control of systematic uncertainties at every step.

## 2.2 Calibrating the distance scale with TRGB stars

As stars reach the tip of the red-giant branch (TRGB) they begin helium burning. The interplay of the triple-alpha reaction, degeneracy and neutrino losses yields a star with a characteristic luminosity at this stage of its life that is only weakly sensitive to its composition and mass.  These are among the brightest stars of a stellar population, and they exist among both old and young populations so we can find them in elliptical and spiral galaxies to large distances.   These stars can provide an independent check on the estimates of the distances to galaxies that host Cepheids (spirals mainly), and they allow us to estimate the distances to elliptical galaxies as well.

HST can measure accurate fluxes of TRGB stars out to about 20 Mpc.  The TRGB luminosity zero-point will soon be calibrated to high precision by Gaia parallaxes.  TRGB distance measurements thus allow a cross-check of the Cepheid distance scale for galaxies that host both classes of stars, and they can calibrate the SNIa luminosity scale for supernovae in early-type host galaxies, allowing an independent check of the "third rung" of the distance ladder that steps from local calibrators to systems in the Hubble flow. Fully exploiting the TRGB distance ladder will ultimately require JWST observations that reach to larger distances and thus a larger population of SNIa hosts. However, observations to ~20 Mpc distances can be done earlier with HST and make better use of the combined HST/JWST resource. Given the high priority of a robust $H_0$ determination, progress on the largely independent approaches towards the distance scale with Cepheids and TRGB stars is highly desirable.

## 2.3 Strong lens time delays

Strong gravitational lens systems with time-variable sources (like AGN and supernovae) provide a measurement of physical distance out to intermediate redshifts, provided the time delays between the multiple images in the system can be measured (typically via ~10-year ground-based optical monitoring campaigns) and the mass distribution in the deflector can be accurately modeled. Time delays in ~10 systems have been measured by the COSMOGRAIL group (e.g. Bonvin et al 2017); in the next decade we can anticipate the LSST providing several hundred measured time delay measurements (Liao et al 2015).

Crucial to the measurement of fundamental physics parameters from time delay lenses is the high resolution imaging of the sources' host galaxies, which appear as "Einstein Rings." Lens mass modeling based on HST Einstein ring images provides enough constraints to enable a 5% precision distance measurement from each system observed. The current precision on the Hubble constant measurement from the H0LiCOW sample of 3 time delay lenses is 3.8%; with 100 systems the aim is to reach sub-percent precision, and provide a vital cosmographic probe that is highly complementary, with very different systematic errors, to distance measurements made in the local or distant Universe (Treu & Marshall 2016). The STRIDES survey should discover 50 or more lensed quasar systems in the Southern hemisphere by 2019, and the COSMOGRAIL team aim to monitor 30 of them. LSST should provide time delays for another few hundred lensed AGN, most of which will be discovered early on in its campaign (c. 2024).

One big open question is whether the high precision mass modeling can be done with sufficient accuracy to yield unbiased cosmological results. The degeneracies inherent to gravitational lens modeling require us to seek out independent mass information about both the lens galaxy and its mass environment (e.g. Suyu et al 2013), as well as focusing on systems with long time delays monitored in extended campaigns (Tewes et al 2013, Tie & Kochanek 2017). HST provides both the Einstein ring constraints and information on the surface brightness of the lens galaxy and its neighbors, enabling 3D mass models to be constructed (Wong et al 2017). A large sample itself would enable a number of internal cross-checks and null tests, in a "statistical strong lensing" regime that we are on the verge of entering. In the end, mitigating lens modeling systematics and achieving accurate results from the ensemble will rely on our being able to employ realistically flexible mass models and correctly marginalizing over the nuisance parameters.

## 2.4 Implications for HST Programs

This report describes a number of ways in which HST observations can provide new insights on fundamental physics. Improved measurement of $H_0$ is our highest priority among these for several reasons.  First, $H_0$ is a core cosmological parameter, and its intrinsic importance is independent of specific cosmological or dark matter scenarios.  Second, $H_0$ is a powerful complement to other cosmological probes, as a diagnostic of dark energy, relativistic species, and other novel physics.  The current tension between "direct $H_0$" and "cosmological $H_0$" values illustrates this power, but a precision measurement of $H_0$ remains valuable even if the tension itself goes away with newer data.  Third, there are clear paths by which HST observations can improve the precision of current $H_0$ measurements, reduce their systematic uncertainties, and achieve cross-checks at interesting levels of precision.

Based on presentations to the committee, a program to obtain Cepheid calibration of ~50 SNIa hosts out to 50 Mpc (including the ~20 already in hand) would require 400-500 orbits. Estimates for TRGB measurements of SNIa hosts to 20 Mpc are somewhat lower, roughly 200 orbits, because of the smaller number of hosts.  HST observations of strong lenses vary by system, but 4-10 orbits is typical, so a campaign of ~25 lenses would again require 100-200 orbits.  It therefore appears that a multi-pronged attack on $H_0$, aimed at achieving 1-2% precision with cross-checks of critical links at a comparable level, could fruitfully use 500-1000 HST orbits distributed over multiple years.  The $H_0$ program may draw on JWST as well as HST, but JWST is an even more highly leveraged resource, so HST observations are preferable where they suffice.

# 3. Dark Matter

The existence of dark matter and dark energy provides perhaps the most compelling evidence for unknown fundamental physics beyond the Standard Model.  While terrestrial accelerators continue to search for signatures of dark matter in collider experiments, HST can provide complementary probes of dark matter properties using a variety of approaches.
The two approaches discussed here are observations of merging galaxy clusters and strong lensing by galaxies.

## 3.1 Merging galaxy clusters

One avenue to constrain dark matter self-interactions utilizes cosmic collisions between systems rich in dark matter.  Massive clusters of galaxies contain dense concentrations of dark matter, and when massive clusters collide, a sensitive probe of the strength of dark matter interactions.  The prototypical example of this is the so-called "Bullet Cluster" 1E 0657-558, which has been used by Markevitch et al. (2004) to bound the elastic scattering SIDM cross-section to $\sigma/m < 1$ cm$^2$/g (subsequent work has led to more robust modeling and modest improvements in the upper limit).  The method used by these authors relies on the fact that ordinary stars in these clusters behave essentially like collisionless particles, so any non-gravitational interactions among dark matter particles could lead to transient spatial offsets between stars and dark matter. These offsets in turn would manifest as offsets between galaxies and local peaks of the lensing mass.

Constraining dark matter interactions using this method requires precise lensing measurements, in both the strong and weak lensing regimes, which HST is well-suited to provide.  Lensing is required not only for precise localization of density peaks, but also for constraining the overall gravitational potential of each system, which is crucial for modeling the merger dynamics and translating bounds on spatial offsets into constraints on interaction cross-sections.  Deep HST imaging over the fields surrounding the Bullet Cluster and similar merging clusters could potentially improve our constraints on SIDM significantly. In addition to lensing data, modeling the merger dynamics may require X-Ray or radio observations of the gas shock and precise spectroscopy for a large number of cluster galaxies to determine line-of-sight velocities.

## 3.2 Substructure in galaxy halos

Small-scale structure in the spatial distribution of dark matter has long been recognized as a sensitive probe of dark matter properties, including its temperature (e.g. Cold vs. Warm DM) or mass scale (as in Fuzzy DM). In particular, the abundance of low-mass substructure in the dark matter halos surrounding typical galaxies is sensitive to a variety of DM physics, and may be used to search for deviations from the standard Cold Dark Matter model. Indeed, the famous "Missing Satellites Problem" in the Milky Way was claimed as motivation for new physics in the dark matter sector. However, since baryonic astrophysics may account for the dearth of faint satellite galaxies in the Local Group, a potentially cleaner probe of small-scale DM structure is gravitational lensing. The underlying goal is to confirm or refute the basic prediction of the CDM scenario, a hierarchy of dark matter halos that extends far below galactic mass scales.

The relative fluxes and positions of multiply imaged systems are sensitive to the presence of small scale DM structure in the lens galaxies. Vegetti et al. (2014) have argued that high-resolution HST imaging of arcs and rings in galaxy-galaxy strong lenses can be used to detect low-mass substructure in the mass distributions of the lensing systems. Since the effects of the lowest-mass substructure on the lensed observables are subtle, and are only detectable with both extremely high resolution and deep multi-band imaging, HST observations are well-suited for this method. Such observations could also complement ongoing searches for lensing substructure by the ALMA array (e.g., Hezaveh et al. 2016). The ongoing Kilo Degree Survey (KiDS), Dark Energy Survey (DES) and Hyper-SuprimeCam (HSC) survey are each expected to discover several hundred strongly lensed galaxies, suggesting that the number of targets suitable for HST imaging may soon increase considerably.

HST observations can also probe the statistics of dark matter substructure using spectroscopy of multiply imaged AGN. Disparate flux ratios of nearby images are a signature of small scale perturbations of the lensing potential (Dalal & Kochanek 2002), but anomalous ratios of the optical continuum can be caused by either subhalo-scale structure or stellar microlensing. Mitigating the microlensing contribution would typically require ~10 year duration ground-based optical monitoring. Alternatively, AGN narrow emission-line regions are too large to be amplified by microlensing, so they allow an application of the substructure test to optical lenses without radio counterparts (Nierenberg et al. 2017). This approach can increase the sample of lenses available for the flux ratio test by a factor of 2-3, with correspondingly improved dark matter constraints.

## 3.3 Implications for HST Programs

Presentations to the committee identified a half-dozen merging cluster systems that would be appropriate candidates for testing self-interacting dark matter theories. Weak lensing observations at the necessary depth require ~50 orbits of imaging per cluster. For strong-lens probes of dark matter substructure, the required orbits per system are typically smaller, and there may be 10-30 systems that would be valuable for such investigations. For both classes of observations, the power of HST will be unsurpassed until the launch of Euclid in the early 2020s and WFIRST in the mid 2020s. HST programs aimed at measuring the properties of dark matter and testing the predictions of CDM could fruitfully use 200-500 orbits distributed over several years. The observations would need to be coupled with detailed modeling of the mass distribution and dynamics of the lens systems.

## 4. "Calibration datasets" for cosmology surveys

Galaxy images from HST's COSMOS field have been invaluable for calibrating photometric redshifts and shear for ground based surveys. Additional color information on the COSMOS galaxies or additional degree-sized fields (with comparable depth to the COSMOS data) would help the calibration of ongoing surveys and ones planned with LSST and Euclid. The gains achievable with a feasible amount of observing time would need to be carefully estimated to justify such an investment. If parallel mode observations can be used to assemble well characterized galaxy images over sufficient sky area, that may be an appealing option to consider.

Lensing shear calibration is primarily done by simulating large numbers of images based on the HST images. The simulated images are then degraded to the specifications of a particular survey and used to calibrate the shear 'responsivity', the correction factor for the rounding effect of the PSF. Statistical errors in Stage 3 surveys like DES are at the few percent level (on the overall amplitude of shear correlations). The COSMOS survey is already 'too small' for Stage 3 surveys in the sense that (a) it is not a fair sample due to clustering along the line of sight, (b) shapes for the full set of galaxies are available in only one filter, and (c) the total number of galaxies is barely sufficient for present needs. Useful advances for the near future can therefore be made by obtaining additional color information on the COSMOS field or adding more area. Euclid's optical survey uses a broad band filter, so it would benefit from HST imaging in even two filters on the full COSMOS field. The challenge for the deep, ground-based imaging from LSST is deblending, so HST data that increases the sampling of the joint (position, flux, size, concentration) distribution of source galaxies will be valuable though it is expensive in observing time. Studies of the ground-space synergy and the calibration needs for Euclid and LSST have been presented in Rhodes et al (2015), Jain et al (2015) and other studies by the projects.

Supernovae are another field where continuing HST observations could add significant value to current and future cosmological surveys through calibrating observations. For relatively nearby supernovae, HST provides access to the rest-frame UV, which is redshifted to visible wavelengths in cosmological surveys of more distant populations. HST also enables supernova observations in the near-IR, where the intrinsic scatter of SNIa luminosities is smaller and uncertainties from dust extinction are reduced. HST supernova samples will be small compared to those from ongoing cosmological surveys, but well-crafted observational programs could provide insights that reduce statistical or systematic errors from ground-based projects such as the Dark Energy Survey and lay essential groundwork for the eventual WFIRST supernova survey.

*Implications for HST Programs*: Observing programs on this topic are not as clearly defined as they are for the $H_0$ and dark matter topics discussed previously. Given the large amounts of observing time that have already been devoted to the COSMOS survey and SNIa observations, it is not clear that further HST programs can achieve a major quantitative advance. However, well designed programs at the tens or hundreds of orbit levels might make an important difference by zeroing in on critical sources of uncertainty, and proposals for such programs should be invited as part of an HST Fundamental Physics theme. This may also be an area where systematic archival investigations of existing HST data or parallel mode observations can achieve valuable results.

# 5. Miscellaneous topics

The precision astrometry and exquisite angular resolution give HST a unique role to probe the properties of individual stars within nearby galaxies. This offers several other avenues to address fundamental physics questions. We describe below an illustrative, and by no means complete, account of such avenues.

The dark-matter dominated dwarf galaxies that surround the Milky Way provide a unique opportunity to probe the properties of dark matter halos and in the process the dark matter itself. Although many of these objects have already been observed with Hubble, HST has a unique capability of to measure proper motions of as many stars as possible in these systems. With proper motions one can get a measurement of the gravitational potential of the system and the structure of the underlying dark matter halo. Follow up programs to the ongoing measurements of Local Group dwarf galaxies have the potential to advance the field. Kallivayalil et al. (2015) outline such a program in a white paper for the Hubble 2020 vision and have begun the observations in a successful HST program (HST Proposal 14734).

Observations of Cepheids and TRGB stars used for the distance ladder would also yield a detailed look at the evolution of stars through the giant branch and the subsequent helium-burning phases. These phases of stellar evolution depend sensitively on the properties of weakly interacting particles, neutrinos in particular but also axions if they exist. By observing these populations in a variety of systems, the effect of environment and different astrophysical phenomena may be mitigated to obtain better constrain the properties of neutrinos and axions.

The presence of supermassive black holes in the centers of most massive galaxies is now well established. HST plays a critical role in providing high resolution observations of the innermost regions of nearby galaxies, as well as characterizing distant AGN. SMBH's allows a unique test of the strong equivalence principle (SEP), again exploiting the high-precision astrometry and the ultraviolet sensitivity of HST. If effects of gravity are different on black holes than on ordinary matter (that is, if the SEP is violated, as expected in recent gravity theories that attempt to explain cosmic acceleration – see Joyce et al (2015) for a review), supermassive black holes will not lie at the precise center of the galaxies that host them. Measurements of the positions of these objects relative to the host galaxies will therefore yield constraints on SEP violations. Furthermore, black hole mergers or even the interactions of black holes with groups of stars can move supermassive black holes away from the centers of galaxies, so the precise measurement of the offsets of these objects also yield estimates of the rate of black hole mergers and the importance of kicks during the merger process.

*Implications for HST Programs:* Observing programs in this category will typically be smaller than those described in previous sections, and they may span a wide range of specific investigations. In principle, these kinds of investigations can be evaluated and prioritized effectively within the standard HST time allocation process. However, the scientific objectives of such observations are tied to testing theories of gravity and particle physics, and if an HST Fundamental Physics program is devised to evaluate the kinds of large observing programs described previously, it would be preferable to evaluate "miscellaneous" proposals within this science theme under the same framework.

# 6. Theory and Simulations

We advocate support for theory and simulations *directly* related to fundamental physics measurements with HST.  Robust cosmological measurements of the precision required to probe the laws of fundamental physics require a deep theoretical understanding of the astrophysical systems used and, even more importantly, validation of the underlying assumptions with detailed simulations of the astrophysics and of observational systematics.  These include (but are not restricted to): modeling the mass distributions for strong-lensing systems used in Hubble parameter measurements; lensing-magnification distributions and supernova-brightness distributions for Hubble parameter measurements; simulations and analytical modeling of particular merging cluster systems for dark matter constraints. In these examples, the theory or simulation effort is closely coupled to particular HST observing programs. We do not advocate HST support for theoretical work that develops new ideas for fundamental physics, work that is already supported by DoE, NSF, NASA ATP, and other programs.

# 7. Recommendation

As the preceding sections make clear, even a quarter-century after launch there are opportunities for HST to make critical contributions to our understanding of dark energy, dark matter, and other aspects of fundamental physics.  However, the kinds of observing programs needed to have a substantive impact on these topics relative to current knowledge are typically demanding in telescope time and technical performance.  The standard system for proposing and selecting Guest Observer programs may not be effective for enabling these contributions, and we recommend an alternative approach.

Specifically, we recommend that the Director allocate observing time over the next three cycles to an "HST Fundamental Physics" program that would be subject to a modified proposal and review process. The criteria for selection in this program should be that a proposal will make a significant advance to the state of knowledge on its topic, or it will provide foundational data that will make an important difference to the expected success of observations with other facilities.  The cases discussed in this report suggest that a total allocation of about 500 orbits/year, or about 1500 orbits total, would be appropriate. This approximate allocation is based on the example cases discussed in Sections 2-6, but any program that falls within the Fundamental Physics category as described in Section 1 could be proposed.

Differences from the standard proposal and review process are:

1. Two-phase proposal submission.  Proposers would submit preliminary proposals that explain the science case and the technical design of their proposed observing program.  A review panel would assess these preliminary proposals and provide feedback in the form of written comments and questions to the science team.  Full proposals could include an Appendix that answers questions raised by the review panel.  Ideally, the *same* panel would review preliminary and full proposals and, to the extent practical, membership in this panel would remain constant or change slowly over the 3-year duration of the program.

2. Multi-year allocations allowed.  Proposals could request all the observing time in a single year, or they could request observations spread over two or three years.  Only a fraction of out-year orbits would be subscribed in the first-year review, so that there is room for other programs to be proposed in subsequent years.

3. Small funding allocations allowed at the preliminary proposal stage. Preliminary proposals could request limited funding for theory or archival research needed to construct a compelling full proposal. This would ensure more careful estimates of observing resources and a better understanding of the modeling risks.

4. Large proposals encouraged.  Meeting the criterion for sufficient impact in Fundamental Physics will typically require ambitious programs that may need hundreds of orbits to be fully realized.  Innovative small proposals may also be competitive, but programs that increment existing data by modest fractions (e.g., increasing the number of observed systems by less than a factor of two) can go through the standard proposal process rather than this modified review.

The full proposal evaluation could be coordinated with that of other HST proposals, or it could be carried out on its own schedule.  In each cycle, the quality and expected impact of observing programs recommended by the Fundamental Physics review panel should be weighed against that of other large HST programs to ensure that the amount of time allocated to the Fundamental Physics effort is appropriate.

We have not recommended a specific observing program in this report because we believe that the best science will emerge from competitive peer review and full engagement by the cosmological community within a special-purpose proposal and review process like that outlined above. The ultimate outcome could be support of one or two very large programs, though we imagine that selection of 3-6 programs is more likely, and a larger number of more diverse investigations is possible as well.

# Acknowledgements
We solicited input from members of diverse research fields for our report. We are especially grateful to Gary Bernstein, Will Dawson, Doug Finkbeiner, Ryan Foley, Wendy Freedman, Mike Jarvis, Alexie Leauthaud, Lucas Macri, Rachel Mandelbaum, Adam Riess, Jason Rhodes, Tim Schrabback, Sherry Suyu, Tommaso Treu, Risa Wechsler and David Wittman.